\documentclass[onecolumn,preprint,noshowkeys,noshowpacs]{revtex4}%
\usepackage{amssymb}
\usepackage{amsmath}
\usepackage{amsfonts}
\usepackage{graphicx}
\usepackage{dcolumn}
\usepackage{bm}%
\providecommand{\U}[1]{\protect\rule{.1in}{.1in}}
\providecommand{\U}[1]{\protect\rule{.1in}{.1in}}

\begin{document}
\title{Effects of alternating interactions and boundary conditions on quantum
entanglement of three-leg Heisenberg ladder }
\author{Qinghui Li$^{1}$, Lizhen Hu$^{2}$, Panpan Zhang$^{1,3}$, Chuanzheng Miao$^{1}%
$, Yuliang Xu$^{1}$, Zhongqiang Liu$^{4}$}
\author{Xiangmu Kong$^{1}$}
\altaffiliation{Corresponding author. E-mail address: kongxm668@163.com (X. Kong).}

\affiliation{$^{1}$School of Physics and Optoelectronic Engineering, Ludong University,
Yantai 264025, China}
\affiliation{$^{2}$School of Physics and Electronic Engineering, Linyi University, Linyi
276000, China}
\affiliation{$^{3}$Department of physics, Beijing Normal University, Beijing 100875, China}
\affiliation{$^{4}$College of Physics and Engineering, Qufu Normal University, Qufu,
273165, China}
\date{\today }

\begin{abstract}
The spin-1/2 three-leg antiferromagnetic Heisenberg spin ladder is studied
under open boundary condition (OBC) and cylinder boundary condition (CBC),
using the density matrix renormalization group and matrix product state
methods, respectively. Specifically, we calculate the energy density,
entanglement entropy, and concurrence while discussing the effects of interleg
interaction $J_{2}$ and the alternating coupling parameter $\gamma$ on these
quantities. It is found that the introduction of $\gamma$ can completely
reverse the concurrence distribution between odd and even bonds. Under CBC,
the generation of the interleg concurrence is inhibited when $\gamma$ $=0$,
and the introduction of $\gamma$ can cause interleg concurrence between chains
1 and 3, in which the behavior is more complicated due to the competition
between CBC and $\gamma$. Additionally, we find that $\gamma$ induces two
types of long-distance entanglement (LDE) in the system under OBC: intraleg
LDE and interleg one. When the system size is sufficiently large, both types
of LDE reach similar strength and stabilize at a constant value. The study
indicates that the three-leg ladder makes it easier to generate LDE compared
with the two-leg system. However, the generation of LDE is inhibited under CBC
which the spin frustration exists. In addition, the calculated results of
energy, entanglement entropy and concurrence all show that there are essential
relations between these quantities and phase transitions of the system.
Further, we predict a phase transition point near $\gamma$ $=0.54$ under OBC.
The present study provides valuable insights into understanding the phase
diagram of this class of systems.

\end{abstract}
\maketitle

\section{Introduction}

As early as 1935, Einstein, Podlsky and Rosen Proposed a special state for two
particles (known as the EPR state), commonly referred to as an entangled
state, which cannot be written as the direct product of states of two
subsystems\cite{1}. Quantum entanglement is the non-local quantum correlation,
indicating that it remains unaffected by the distance between the two
subsystems and persists regardless of how far apart they are. This concept is
vital in quantum mechanics and plays a crucial role in various fields,
including condensed matter physics and quantum information science, such as
quantum computing, quantum communication, quantum cryptography,
etc.\cite{2,3,4,5,6,7,8,9}.

Back in 2002, Osterloh et al. investigated the quantum entanglement in
one-dimensional XY and Ising models under external magnetic fields. They
analyzed the ground state wave function in the critical region and uncovered
the connection between quantum entanglement and quantum phase
transitions\cite{10,10(1)}. Following this pioneering work, the study of
quantum entanglement in spin systems has garnered increasing
interest\cite{11,12,13,14}. As is well-known to all, one-dimensional spin
systems have computational advantages and can be solved exactly or treated
with the renormalization group method. Significant progress has been made in
elucidating the relationship between quantum entanglement and phase
transitions in these one-dimensional spin
systems\cite{16,19(1),19(2),19(4),19(6),19(7)}.

In the other aspect, spin ladders play a crucial role in explaining the
crossover behaviour between one-dimensional and two-dimensional Heisenberg
antiferromagnets. These systems can model a class of materials, including
Sr$_{n-1}$Cu$_{n+1}$O$_{2n}$, La$_{6}$Ca$_{8}$Cu$_{24}$O$_{41}$ and Cu$_{2}%
$(C$_{5}$H$_{12}$N$_{2}$)$_{2}$Cl$_{4}$, among others\cite{20,21,22}. In 1996,
Dagotto et al. demonstrated that Heisenberg spin ladders with an even number
of legs feature energy gaps and exhibit short-range correlations\cite{22(1)}.
In contrast, ladders with an odd number of legs lack energy gaps and instead
display power-law correlations\cite{49}, which is similar to the corresponding
spin chains. These theoretical findings have been confirmed in actual
materials such as (VO)$_{2}$P$_{2}$O$_{7}$ and the cuprate series\cite{23}.
Moreover, both experimental and theoretical investigations have revealed
fascinating crossover behaviours, such as significant magnetic field effects
and dynamic properties\cite{23,24,25,26,27}. Research shows that
antiferromagnetic systems, when there is spin frustration and strong quantum
fluctuations, manifest unique phenomena\cite{28}. Recently, Almeida et al.
used the density matrix renormalization group (DMRG) method and a hard-core
boson mapping technique to investigate the phase diagram of the frustrated
Heisenberg ladder under a magnetic field. They gave the conditions for the
exchange coupling parameters when there are first-order phase transition lines
and bicritical points in the phase diagram\cite{29}.

In 2003, Bose designed a mechanism for short-range quantum communication using
spin chains as quantum channels\cite{30}. Subsequently, it is proposed that a
strong and stable entanglement between two spins that are far away and do not
directly interact could be used for long-distance quantum communication. As we
all know, in a general short-range interacting system, the pairwise
entanglement decays rapidly as the distance between the two spins
increases\cite{5,32}. However, Venuti et al. discovered that the XXX spin
chain with alternating interactions exhibits long-distance entanglement (LDE)
under open boundary conditions (OBC)\cite{33}. In the experiment, Sahling et
al. confirmed the presence of LDE in antiferromagnetic spin chains with
alternating interactions for the first time in 2015\cite{34}. This finding has
sparked significant interest in the application of alternating interaction
spin models within the field of quantum
information\cite{35,36,37,38,39,40,41,42}. In 2022, using the DMRG method, we
studied the two-leg antiferromagnetic Heisenberg spin system and found that
the alternating interactions are conducive to the generation and enhancement
of LDE, while the anisotropic interaction inhibits LDE\cite{43}.

For the three-leg Heisenberg spin ladder, in the last 20 years, Azzouz et al.
calculated the phase diagram and magnetic properties; however, there are few
studies on quantum entanglement and quantum correlation\cite{44,45,46,47,48}.
In 1996, Frischmuth investigated the magnetic susceptibility and entropy of
one-, three-, and five-leg Heisenberg spin ladders under uniform and staggered
interactions, respectively, by using the Monte Carlo cycle algorithm and exact
diagonalization method\cite{49}. In 2002, Wang studied the ground-state phase
diagram of a spin-1/2 frustrated three-leg antiferromagnetic Heisenberg
ladder, a symmetric doublet phase and a quartet one were found in the system
by the DMRG method\cite{50}. More recently, in 2016, Wang applied the bond
mean field method to study the magnetism of a three-leg antiferromagnetic
spin-1/2 Heisenberg ladder. They provided the mean-field bond parameters and
computed the concurrence values, thereby substantiating the occurrence of
phase transitions\cite{51}.

In this paper, we investigate the three-leg antiferromagnetic spin-1/2
Heisenberg ladder through the DMRG method, using matrix product states (MPS)
as described in reference\cite{52}. We quantify the dependence of the energy
density, entanglement entropy, and concurrence on the interleg interaction
$J_{2}$ and the alternating coupling parameter $\gamma$ under OBC and cylinder
boundary condition (CBC), respectively. This paper is organized as follows: In
Sec. \ref{sec2}, we present the Hamiltonian of the system and introduce the
DMRG algorithm, calculating the energy of the system when $\gamma=0$; In Sec.
\ref{sec3}, we calculate the nearest-neighbour concurrence under the OBC and
CBC for $\gamma=0$. In Sec. \ref{sec4}, the effects of $\gamma$ and CBC on the
entanglement of the system are discussed. Sec. \ref{sec5} is the conclusion.

\section{Model and energy density\label{sec2}}

Consider a three-leg Heisenberg spin ladder with alternating coupling
parameter $\gamma$ (the structure diagram is shown in Fig. 1), and its
Hamiltonian can be expressed as%
\begin{equation}
H=\frac{1}{4}\left[  \underset{k=1}{\overset{3}{\sum}}\underset{i=1}%
{\overset{L-1}{\sum}}J_{1}\left(  1+\left(  -1\right)  ^{i+k-1}\gamma\right)
\overset{\rightharpoonup}{\sigma}_{i,k}\cdot\overset{\rightharpoonup}{\sigma
}_{i+1,k}+\underset{k=1}{\overset{n}{\sum}}\underset{i=1}{\overset{L}{\sum}%
}J_{2}\left(  \overset{\rightharpoonup}{\sigma}_{i,k}\cdot\overset
{\rightharpoonup}{\sigma}_{i,k+1}\right)  \right]  , \label{1}%
\end{equation}
where $\overset{\rightharpoonup}{\sigma}_{i,k}=\sigma_{i,k}^{x}\overset
{\rightharpoonup}{i}+$ $\sigma_{i,k}^{y}\overset{\rightharpoonup}{j}+$
$\sigma_{i,k}^{z}\overset{\rightharpoonup}{k}$, $\sigma_{i,k}^{\alpha}$
($\alpha=x,y,z$) are the Pauli operators on the $i$-th site in the chain-$k$,
and the corresponding matrix form in the $\sigma^{z}$ representation are%
\begin{equation}
\sigma_{i,k}^{x}=\binom{0\text{ }1}{1\text{ }0}\text{ }\sigma_{i,k}^{y}%
=\binom{0\text{ -}I}{I\text{ }0},\text{ }\sigma_{i,k}^{z}=\binom{1\text{ \ }%
0}{0\text{ -}1}, \label{2}%
\end{equation}
in which $I$ is the imaginary unit. $L$ is the length of a single chain;
$J_{1}>0$ and $J_{2}>0$ represent the antiferromagnetic interactions of
nearest-neighbour spins along legs (intraleg, $x$ direction) and rungs
(interleg, $y$ direction), respectively. When $n=2$, the system is in OBC;
When $n=3$, $\sigma_{i,4}^{\alpha}=\sigma_{i,1}^{\alpha}$, there is an
interaction $J_{2}$ between the nearest-neighbour spins of the first and third
chains, corresponding to CBC in the $y$ direction of the system, in which the
spin frustration exists. $\gamma=0$ corresponds to the special case of uniform
interactions, and $0<\gamma<1$ corresponds to the case of alternating
coupling. $J_{1}(1-\gamma)$ and $J_{1}(1+\gamma)$ describe the "weak bond" and
"strong bond", which are shown by the dashed and solid red lines in Fig. 1, respectively.

In this study, we use the DMRG method to calculate the energy and entanglement
of the system. It is well known that in this method, any target state
$\left\vert \psi\right\rangle $ can be represented as the superposition form
of two subsystems $\left\vert \psi\right\rangle =\underset{i,j}{%
%TCIMACRO{\dsum }%
%BeginExpansion
{\displaystyle\sum}
%EndExpansion
}c_{i,j}\left\vert i\right\rangle _{A}\left\vert j\right\rangle _{B}$. By
applying singular value decomposition, the state can be represented as
$\left\vert \psi\right\rangle =\underset{a}{%
%TCIMACRO{\dsum }%
%BeginExpansion
{\displaystyle\sum}
%EndExpansion
}\omega_{\alpha}\left\vert \mu_{\alpha}\right\rangle _{A}\left\vert
\mu_{\alpha}\right\rangle _{B}$, where $\left\vert \omega_{\alpha}\right\vert
^{2}$ indicates the proportion of the $\left\vert \mu_{\alpha}\right\rangle $
in $\left\vert \psi\right\rangle $. By preserving at most $m$ maximum
eigenvalues $\omega_{\alpha}$ and their corresponding eigenvector $\left\vert
\mu_{\alpha}\right\rangle $, the target state can be represented by the
truncated state $\left\vert \widetilde{\psi}\right\rangle $, with the
truncation error $\left\vert \left\vert \psi\right\rangle -\left\vert
\widetilde{\psi}\right\rangle \right\vert ^{2}$. The ground state is set as
the target state, and a maximum of $m=400$ states are reserved, which is
enough to keep the truncation error $\sim$10$^{-9}$. In order to avoid the
program cannot accurately identify the system when it is in ground state
degeneracy, perturbation $p\left(  \underset{i}{%
%TCIMACRO{\tsum }%
%BeginExpansion
{\textstyle\sum}
%EndExpansion
}\sigma_{i}\right)  ^{2}$, $p=0.001$, is introduced into the Hamiltonian.%

%TCIMACRO{\FRAME{ftbpFU}{5.1223in}{1.5454in}{0pt}{\Qcb{Schematic diagram of the
%three-leg alternately coupled spin ladder. The nearest neighbor interactions
%in the $x$ direction are alternately strong and weak.}}{}{fig1.eps}%
%width 5.1223in;  height 1.5454in;  depth 0pt;  original-width 3.2353in;
%original-height 0.8674in;  cropleft "0";  croptop "1";  cropright "1";
%cropbottom "0";  filename 'Fig1.eps';file-properties "XNPEU";}}}%
%BeginExpansion
\begin{figure}
[ptb]
\begin{center}
\includegraphics[
height=1.5454in,
width=5.1223in
]%
{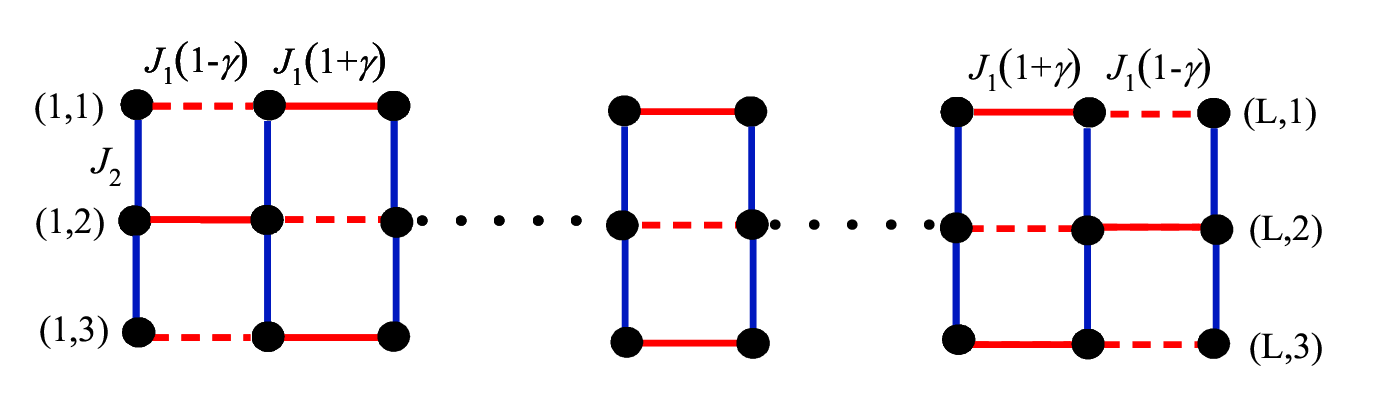}%
\caption{Schematic diagram of the three-leg alternately coupled spin ladder.
The nearest neighbor interactions in the $x$ direction are alternately strong
and weak.}%
\end{center}
\end{figure}
%EndExpansion

In this section, we initially investigate the case of $\gamma=0$ under the OBC
and CBC (during the calculation that $J_{1}=1$). The energy density of the
ground state, $e_{0}=E_{0}/N$, as well as the energy densities for the first
and second excited states ($e_{1\text{ }}$and $e_{2}$), are investigated.
Additionally, the energy gaps $\Delta e_{1}=$ $e_{1}-e_{0}$ and $\Delta
e_{2}=$ $e_{2}-e_{0}$ are examined as functions of the parameters $L$ and
$J_{2}$, which are shown in Fig. 2. The effects of boundary condition, system
size $L$ and interleg interaction $J_{2}$ on them are discussed.%

%TCIMACRO{\FRAME{ftbpFU}{6.0485in}{2.8063in}{0pt}{\Qcb{(a) The variations of
%the energy density of ground state $e_{0}$ with $L$ for certain values of
%$J_{2}$, and $e_{0}$ gradually converges to a stable value with the increase
%of $L$. (b) The variations of $\Delta e_{1}=$ $e_{1}-e_{0}$ with $L$ for
%certain $J_{2}$, showing a trend towards zero as $L$ increases. The inset
%gives the variations of $\Delta e_{1}$ and $\Delta e_{2}$ $=$ $e_{2}-e_{0}$
%with $L$ for $J_{2}=0.5$ under the OBC. The full and half-filled symbols
%correspond to the systems under the OBC and CBC, respectively.}}{}%
%{fig2.eps}{\special{ language "Scientific Word";  type "GRAPHIC";
%width 6.0485in;  height 2.8063in;  depth 0pt;  original-width 17.4208in;
%original-height 7.9649in;  cropleft "0";  croptop "1";  cropright "1";
%cropbottom "0";  filename 'Fig2.eps';file-properties "XNPEU";}}}%
%BeginExpansion
\begin{figure}
[ptb]
\begin{center}
\includegraphics[
height=2.8063in,
width=6.0485in
]%
{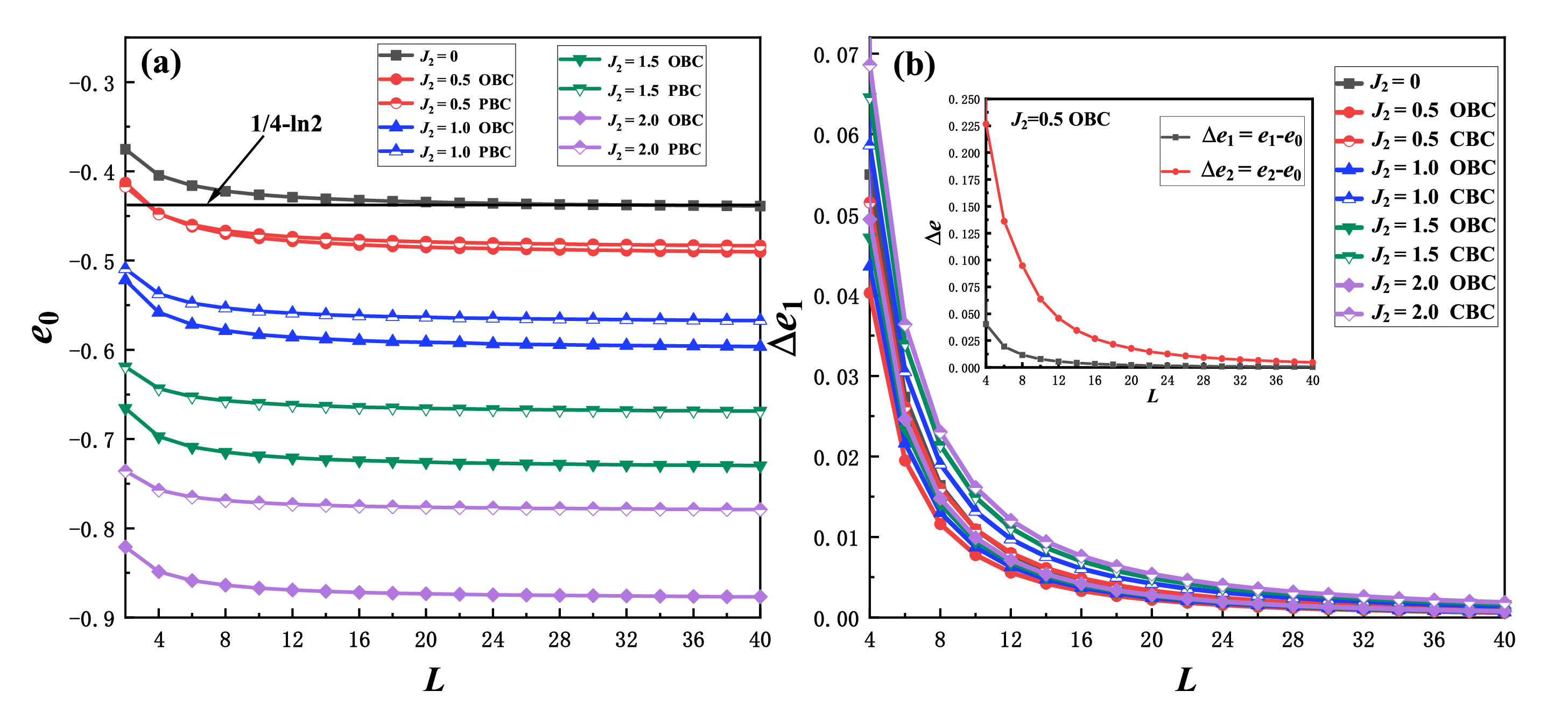}%
\caption{(a) The variations of the energy density of ground state $e_{0}$ with
$L$ for certain values of $J_{2}$, and $e_{0}$ gradually converges to a stable
value with the increase of $L$. (b) The variations of $\Delta e_{1}=$
$e_{1}-e_{0}$ with $L$ for certain $J_{2}$, showing a trend towards zero as
$L$ increases. The inset gives the variations of $\Delta e_{1}$ and $\Delta
e_{2}$ $=$ $e_{2}-e_{0}$ with $L$ for $J_{2}=0.5$ under the OBC. The full and
half-filled symbols correspond to the systems under the OBC and CBC,
respectively.}%
\end{center}
\end{figure}
%EndExpansion

Figure 2(a) illustrates the relation between $e_{0}$ and $L$ for several
values of $J_{2}$. As $L$ increases, $e_{0}$ gradually decreases and converges
to a stable value. Notably, when $L>24$, $e_{0}$ remains constant, indicating
that it is not sensitive to the system size. Therefore, we will use $L=24$ for
the subsequent calculations. By comparing the variations of $e_{0}$ under the
two boundary conditions with $J_{2}$, we find that due to spin frustration in
the system under CBC, the ground state energy density under OBC is lower than
that under CBC, and their difference increases with the increase of $J_{2}$.
Additionally, the $e_{0}$ gradually decreases with $J_{2}$ increases, which
indicates that $J_{2}$ exerts an inhibition effect on the energy density. In
the special case of $J_{2}=0$, $e_{0}$ gradually converges to 1/4 - ln 2 as
$L$ increases, which is consistent with the result given by Bethe Ansatz
method\cite{53}.

Figure 2(b) displays the variations of $\Delta e_{1}$ with $L$ for certain
$J_{2}$. The value of $J_{2}$ has little impact on $\Delta e_{1}$, and $\Delta
e_{1}$ gradually decreases to zero with the increase of $L$, which is
consistent with the result that odd-leg ladders have no energy gap\cite{49}.
The difference between the energy densities $e_{2}$ and $e_{0}$ is further
studied, and the results are presented in the inset of Fig. 2(b). As an
example, under OBC, when $J_{2}=0.5$, both $\Delta e_{1}$ and $\Delta e_{2}$
gradually decrease to zero as $L$ increases. Through calculations of energy
density, we find that the degeneracies of both$\ e_{1}$ and the $e_{2}$ are
three, indicating that the ground state and first and second excited states
are degenerate under the thermodynamic limit, so the degeneracy of the
system's ground state is seven.

\section{Concurrence in Heisenberg spin ladder\label{sec3}}

In the previous section, we calculate the energy density of the system. Next,
we investigate the entanglement properties when $\gamma=0$, the
nearest-neighbor pairwise entanglement in the ground state is calculated, and
the influence of $J_{2}$ and boundary condition on the entanglement properties
is studied. The pairwise entanglement can be measured by concurrence $C_{i,j}%
$, which represents the entanglement between the $i$-th and $j$-th sites in a
many-body system. The $C_{i,j}$ can be calculated using the
expression\cite{54},
\begin{equation}
C_{i,j}=\max\left\{  \sqrt{\lambda_{4}}-\sqrt{\lambda_{3}}-\sqrt{\lambda_{2}%
}-\sqrt{\lambda_{1}},0\right\}  \text{,} \label{3}%
\end{equation}
where $\lambda_{k}$ $\left(  k=1,2,3,4\right)  $ are the non-negative
eigenvalues of $\rho_{i,j}\overset{\sim}{\rho}_{i,j}$with increasing order,
and the reduced density matrix $\rho_{i,j}$ for the two spins $i$ and $j$ can
be obtained by tracing out other spins $\rho_{i,j}=$Tr$_{\sigma\neq\sigma
_{i},\sigma_{i}}\left\vert \psi\right\rangle \left\langle \psi\right\vert $,
$\overset{\sim}{\rho}_{i,j}=\left(  \sigma_{i}^{y}\otimes\sigma_{j}%
^{y}\right)  \rho_{i,j}^{\ast}\left(  \sigma_{i}^{y}\otimes\sigma_{j}%
^{y}\right)  $ is the spin-flip density matrix operator of $\rho_{i,j}$,
$\rho_{i,j}^{\ast}$ is the complex conjugate of $\rho_{i,j}$.

\subsection{Intraleg entanglement\label{sec3.1}}

\subsubsection{Spatial distributions of concurrence}

We study the spatial distributions of intraleg concurrence $C_{i\text{-}i+1}$
(where $C_{i\text{-}i+1}\equiv C_{i,i+1}$ are the concurrence of bond between
sites $i$ and $i+1$, $i=$ odd (even) corresponding odd (even) bonds) under two
boundary conditions. In the case of OBC, each site in chain-$k$ ($k=1$ and
$3)$ is connected by a single interaction $J_{2}$; however, each site in
chain-$k$ ($k=2$) is connected by two interactions. Due to the different spin
interactions present in each chain, the concurrence behaviour observed in
$k=2$ is different from that in $k=1(3)$. Under CBC, each site in chain-$k$
($k=1,2,3$) has the same number of interactions $J_{2}$ connecting each site,
leading all chains to display similar entanglement properties.%

%TCIMACRO{\FRAME{ftbpFU}{6.6331in}{2.0141in}{0pt}{\Qcb{Spatial distributions of
%intraleg concurrence with $L=24$. (a) and (b) show the distributions in
%$k=1(3)$ and $2$ under OBC, respectively. (c) shows the distributions under
%CBC. There exist the phenomena of entanglement separation between odd and even
%bonds. The full-filled symbols represent the concurrence on odd bonds, and the
%half-filled symbols represent the concurrence on even bonds.}}{}%
%{fig3.eps}{\special{ language "Scientific Word";  type "GRAPHIC";
%width 6.6331in;  height 2.0141in;  depth 0pt;  original-width 26.4235in;
%original-height 7.9442in;  cropleft "0";  croptop "1";  cropright "1";
%cropbottom "0";  filename 'Fig3.eps';file-properties "XNPEU";}}}%
%BeginExpansion
\begin{figure}
[ptb]
\begin{center}
\includegraphics[
height=2.0141in,
width=6.6331in
]%
{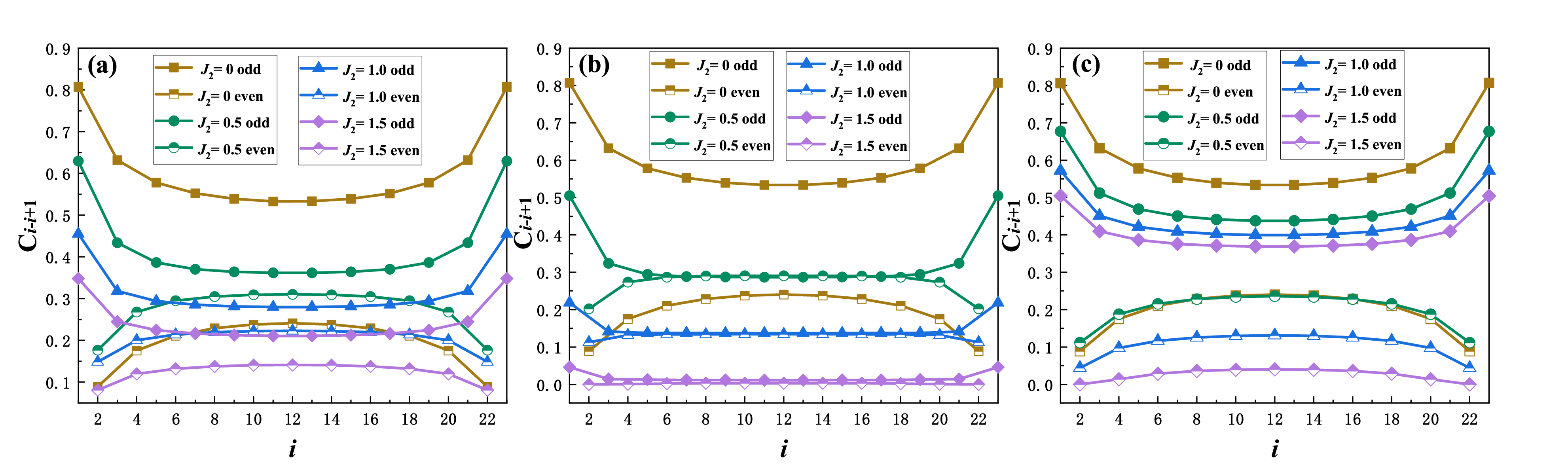}%
\caption{Spatial distributions of intraleg concurrence with $L=24$. (a) and
(b) show the distributions in $k=1(3)$ and $2$ under OBC, respectively. (c)
shows the distributions under CBC. There exist the phenomena of entanglement
separation between odd and even bonds. The full-filled symbols represent the
concurrence on odd bonds, and the half-filled symbols represent the
concurrence on even bonds.}%
\end{center}
\end{figure}
%EndExpansion

Figure 3 shows the spatial distributions of intraleg concurrence in $k=1(3)$
and $2$ under OBC and CBC with $L=24$, respectively. The results indicate that
when $J_{2}$ takes a certain value, the concurrence of odd bonds is greater
than that of even bonds, leading to a phenomenon known as entanglement
separation between odd and even bonds. Despite the uniform interactions among
all spins in the chain, the open boundary condition in the x-direction results
in an alternating dimerization of concurrence, which enhances the entanglement
of odd bonds located near both ends of the chain. Similar phenomena have also
been reported in previous literature that studied the spin-1/2 Heisenberg
antiferromagnet and XXZ chains\cite{55,56}.

Although the spatial distributions of intraleg concurrence exhibit a similar
trend under two boundary conditions, the difference in concurrence between odd
and even bonds varies significantly for certain values of $J_{2}$. Taking
$J_{2}=1.5$ as an example, the value of the difference of concurrence between
odd and even bonds in the middle of the chain is approximately $0.07$ in
$k=1$, while it is nearly zero in $k=2$ under OBC. It is worth noting that in
the case of CBC, this difference becomes more pronounced, indicating that the
effect of $J_{2}$ on concurrence varies under the two boundary conditions.%

%TCIMACRO{\FRAME{ftbpFU}{5.6792in}{1.695in}{0pt}{\Qcb{Schematic diagrams of the
%bonds studied in the system with $L=24$. (a) Four intraleg bonds: 1-2, 2-3,
%11-12, 12-13. (b) Four interleg bonds: 1, 2, 11, 12. }}{}{fig4.eps}%
%width 5.6792in;  height 1.695in;  depth 0pt;  original-width 2.5815in;
%original-height 0.9919in;  cropleft "0";  croptop "1";  cropright "0.9946";
%cropbottom "0";  filename 'Fig4.eps';file-properties "XNPEU";}}}%
%BeginExpansion
\begin{figure}
[ptb]
\begin{center}
\includegraphics[
trim=0.000000in 0.000000in 0.013940in 0.000000in,
height=1.695in,
width=5.6792in
]%
{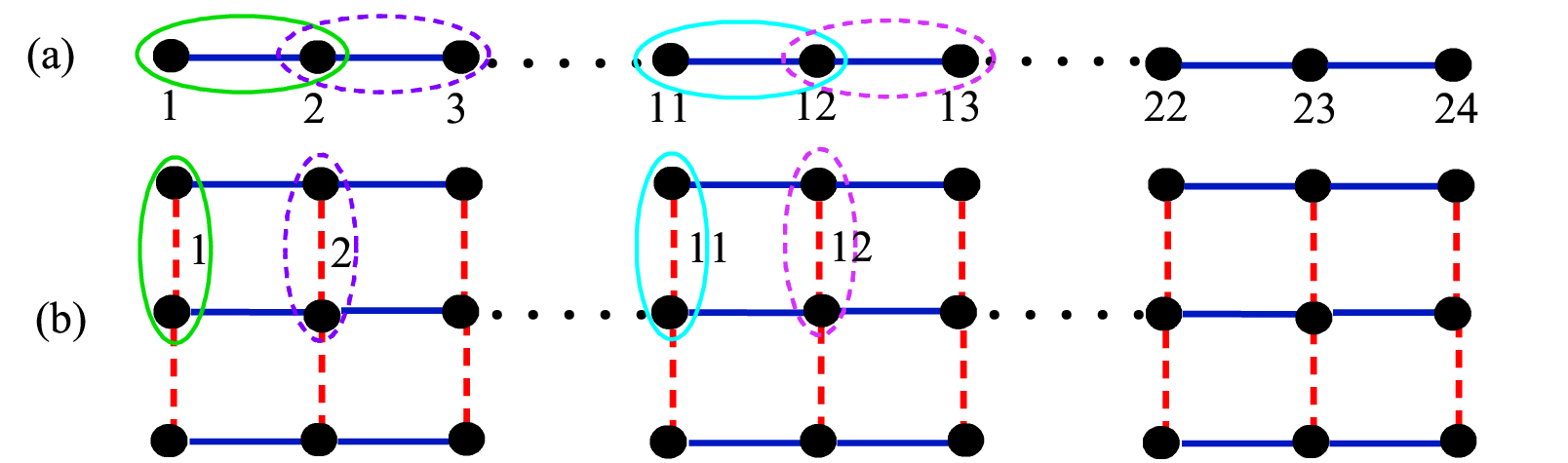}%
\caption{Schematic diagrams of the bonds studied in the system with $L=24$.
(a) Four intraleg bonds: 1-2, 2-3, 11-12, 12-13. (b) Four interleg bonds: 1,
2, 11, 12. }%
\end{center}
\end{figure}
%EndExpansion

\subsubsection{The effect of $J_{2}$ on the concurrence}%

%TCIMACRO{\FRAME{ftbpFU}{5.7354in}{5.0176in}{0pt}{\Qcb{Relation between
%concurrence and $J_{2}$ among nearest-neighbor spins in the chain, $L=24$: (a)
%and (b) are the variations of concurrence with $J_{2}$ (OBC) of chain-1 and
%chain-2, respectively. There are two different trends of odd and even bond
%entanglement. (c) is the variation of concurrence with $J_{2}$ (CBC). (d)
%Variations of the difference between the concurrence of nearest-neighbor bonds
%with $J_{2}$.}}{}{fig5.eps}{\special{ language "Scientific Word";
%type "GRAPHIC";  maintain-aspect-ratio TRUE;  display "USEDEF";
%valid_file "F";  width 5.7354in;  height 5.0176in;  depth 0pt;
%original-width 12.4187in;  original-height 10.9823in;  cropleft "0";
%filename 'Fig5.eps';file-properties "XNPEU";}}}%
%BeginExpansion
\begin{figure}
[ptb]
\begin{center}
\includegraphics[
height=5.0176in,
width=5.7354in
]%
{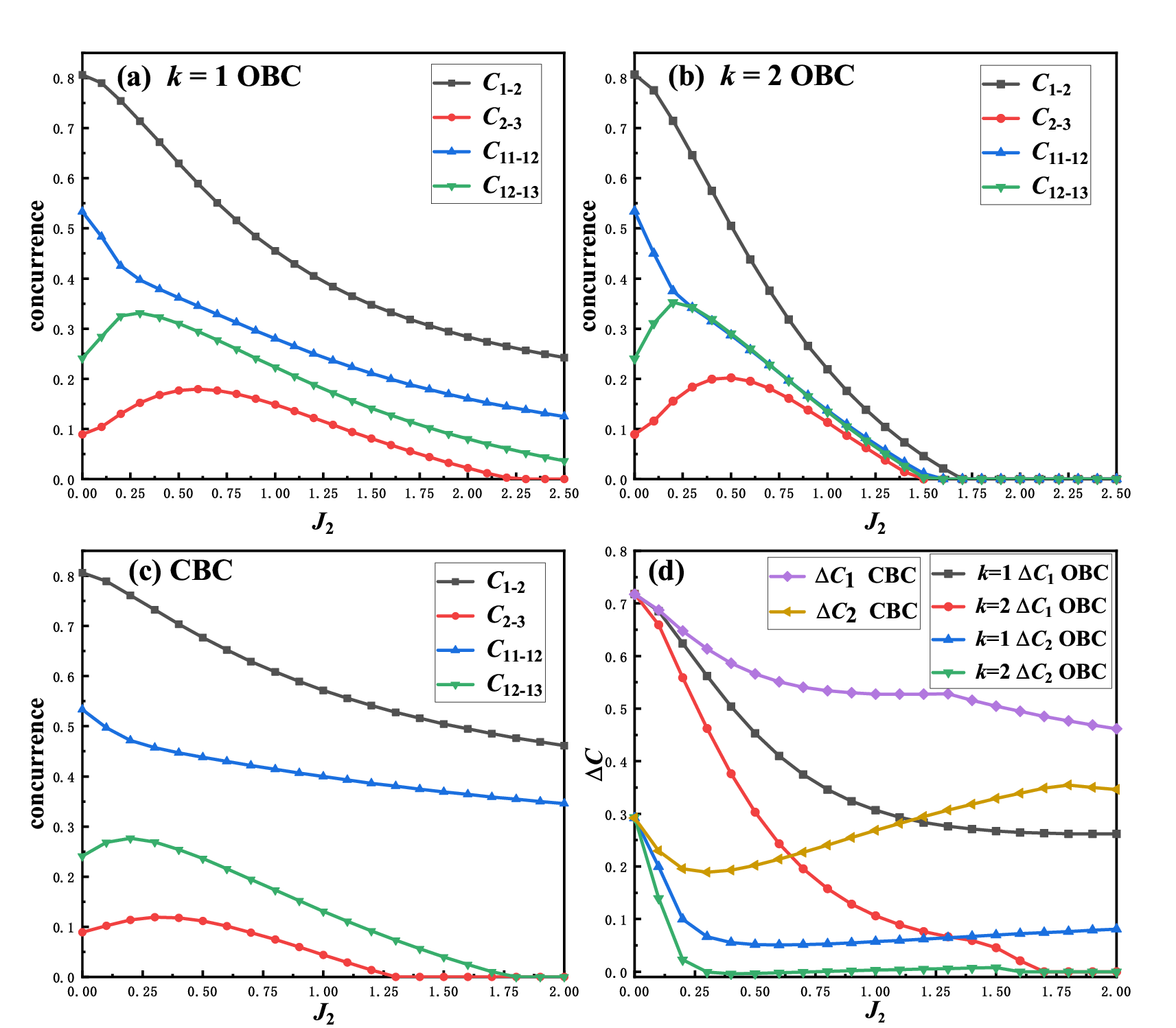}%
\caption{Relation between concurrence and $J_{2}$ among nearest-neighbor spins
in the chain, $L=24$: (a) and (b) are the variations of concurrence with
$J_{2}$ (OBC) of chain-1 and chain-2, respectively. There are two different
trends of odd and even bond entanglement. (c) is the variation of concurrence
with $J_{2}$ (CBC). (d) Variations of the difference between the concurrence
of nearest-neighbor bonds with $J_{2}$.}%
\end{center}
\end{figure}
%EndExpansion

To further investigate the effects of $J_{2}$ and boundary conditions on
intraleg entanglement, we calculate the concurrence of four typical bonds
$C_{1\text{-}2}$, $C_{2\text{-}3}$, $C_{11\text{-}12}$ and $C_{12\text{-}13}$
(see Fig. 4(a)). It can be observed from Fig.5 that with the increase of
$J_{2}$, the concurrence of odd bonds ($C_{1\text{-}2}$ and $C_{11\text{-}12}%
$) gradually decreases, i.e., $J_{2}$ exerts a suppressive effect on
concurrence within odd bonds. Meanwhile, the concurrence of the even bonds
($C_{2\text{-}3}$ and $C_{12\text{-}13}$) initially increases and then
subsequently decreases. A comparison of the concurrence variations with
$J_{2}$ reveals that the concurrence of chain-2 rapidly decreases to zero as
$J_{2}$ increases, indicating that the inhibiting effect of $J_{2}$ on
concurrence in chain-2 is more pronounced than in chain-1. Additionally, we
note that the strength of the concurrence in the odd (even) bonds differs
between the two chains, and as $J_{2}$ increases, $C_{1\text{-}2}$ and
$C_{11\text{-}12}$ of chain-1 consistently remain greater than those in
chain-2. $C_{2\text{-}3}$ and $C_{12\text{-}13}$ of chain-2 is larger than
those in chain-1 when $J_{2}<0.75$ and $0.4$, respectively, as detailed in
Figs. 5(a) and 5(b). The boundary conditions do not affect the variation trend
of bond concurrence with $J_{2}$, but the difference between $C_{11\text{-}%
12}$ and $C_{12\text{-}13}$ under CBC is significantly larger than that under
OBC (see Fig. 5(c)).

\subsubsection{Dimerization of the concurrence}

To delve into the impact of boundary conditions on the dimerization of
concurrence, we use the difference in concurrence between nearest-neighbour
bonds, $\Delta C_{1}=C_{1\text{-}2}-C_{2\text{-}3}$ and $\Delta C_{2}%
=C_{11\text{-}12}-C_{12\text{-}13}$, as measures of the concurrence
dimerization strength. As depicted in Fig. 5(d), as $J_{2}$ increases, $\Delta
C_{1}$ under OBC and CBC gradually decreases. However, the reduction in
$\Delta C_{1}$ under CBC is significantly smaller compared to that under OBC,
indicating that $J_{2}$ plays a suppressive effect on the dimerization effect
of the bonds at both ends of the chain. Conversely, the frustration caused by
CBC plays a promoting effect on the dimerization effect. Different from
$\Delta C_{1}$, $\Delta C_{2}$ initially decreases before increasing as
$J_{2}$ increases, with this trend being particularly noticeable under CBC.
This suggests that the inhibitory effect of $J_{2}$ on the dimerization of the
central part of the chain transitions from strong to weak. Considering the
CBC, the system demonstrates a phenomenon known as spin frustration. The
competition between the frustration and $J_{2}$ reaches a balance when
$J_{2}\approx0.3$, and the promotion effect of CBC on the dimerization effect
is stronger than the inhibition effect of $J_{2}$ when $J_{2}>0.3$.

\subsection{Interleg entanglement\label{sec3.2}}%

%TCIMACRO{\FRAME{ftbpFU}{6.359in}{2.8055in}{0pt}{\Qcb{Interleg concurrence,
%$L=24$: (a) Spatial distributions of the interleg concurrence. (b) Concurrence
%of four representative interleg bonds varies with $J_{2}$, which first appears
%at $J_{2}=0.2$. The inset shows the derivative of $C_{11}$ with respect to
%$J_{2}$, and there is a maximum value at $J_{2}=0.6$.}}{}{fig6.eps}%
%width 6.359in;  height 2.8055in;  depth 0pt;  original-width 20.4234in;
%original-height 8.9629in;  cropleft "0";  croptop "1";  cropright "1";
%cropbottom "0";  filename 'Fig6.eps';file-properties "XNPEU";}}}%
%BeginExpansion
\begin{figure}
[ptb]
\begin{center}
\includegraphics[
height=2.8055in,
width=6.359in
]%
{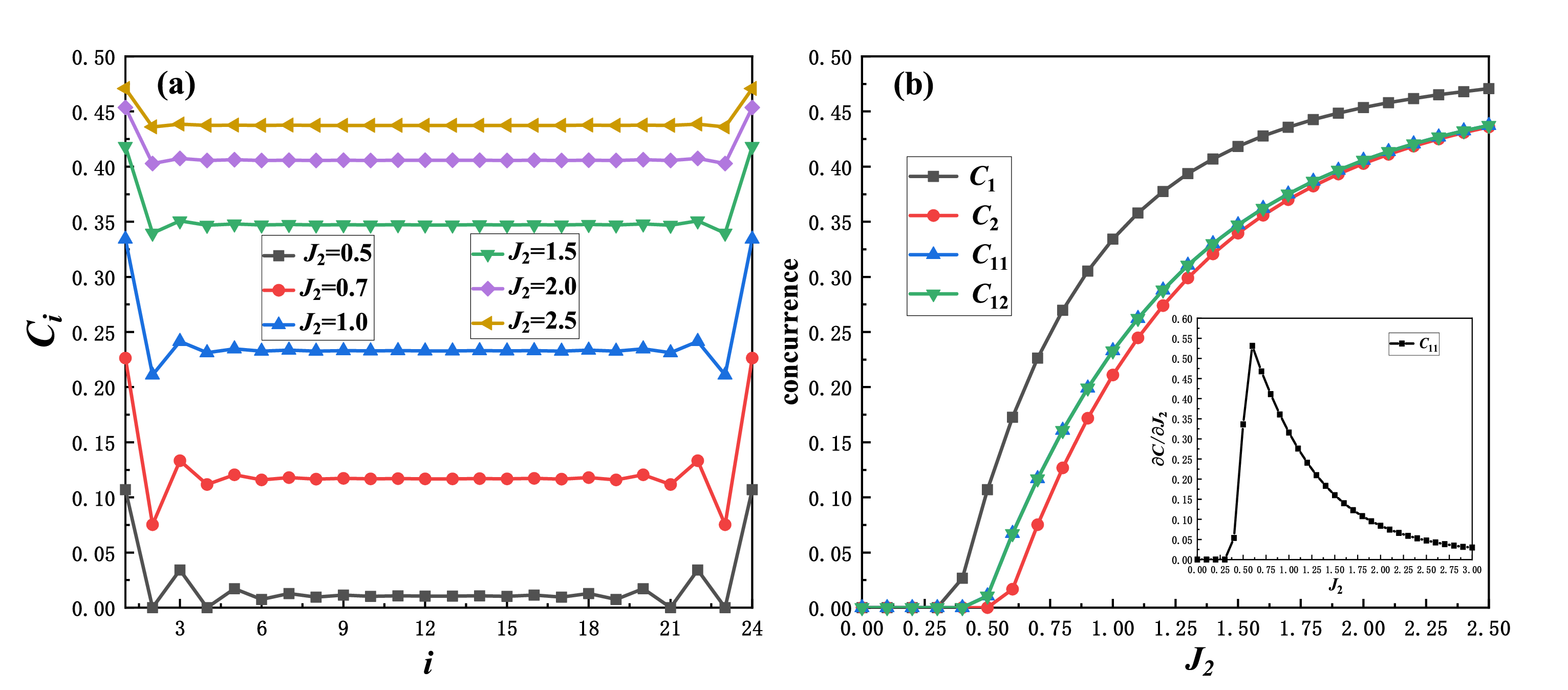}%
\caption{Interleg concurrence, $L=24$: (a) Spatial distributions of the
interleg concurrence. (b) Concurrence of four representative interleg bonds
varies with $J_{2}$, which first appears at $J_{2}=0.2$. The inset shows the
derivative of $C_{11}$ with respect to $J_{2}$, and there is a maximum value
at $J_{2}=0.6$.}%
\end{center}
\end{figure}
%EndExpansion

In this subsection, we calculate the spatial distribution of interleg
concurrence $C_{i}$ and the concurrence in four bonds $C_{1}$, $C_{2}$,
$C_{11}$ and $C_{12}$ (corresponding positions in the system with $L=24$ are
shown in Fig. 4(b)) under OBC (see Fig. 6). Figure 6(a) shows the spatial
distribution when $J_{2}$ takes several values. Obviously, the concurrence
values at both ends are the largest, and the remaining concurrence values are
basically in a uniform distribution, and the larger $J_{2}$ is, the larger the
interleg concurrence is. In Fig. 6(b), when $J_{2}$ is small, the interleg
concurrence does not exist; when $J_{2}$ closes to $J_{2a}=0.2$, $C_{1}$
appears first and then gradually increases. $C_{2}$, $C_{11}$ and $C_{12}$
have similar phenomena to $C_{1}$, but the values of $J_{2a}^{{}}$ are
different. The inset is the curve of the derivative of $C_{11}$ with respect
to $J_{2}$, in which there is a maximum value of $C_{11}$ at $J_{2}=0.6$.

Additionally, we study the case of CBC and discover that $C_{i}$ always
remains at zero, i.e., they are in a dead state as $J_{2}$ increases. It can
be verified that, in this case, the spin frustration phenomenon exists, which
inhibits the generation of interleg concurrence.

\section{Heisenberg ladder with alternating couplings\label{sec4}}

The above section, we also calculate the LDE in the case of $\gamma=0$. The
findings indicate that LDE (e.g., concurrence between sites 1 and $L$ in
chain-1) is zero in this case. Previous study has shown that the introduction
of $\gamma$ is conducive to the generation of LDE, which can introduce more
abundant entanglement phenomena into the system\cite{43}. In this section,
considering OBC, we investigate the influence of $\gamma$ on the system's
energy, entanglement entropy and concurrence, especially LDE, and further
discuss the relation between quantum entanglement and quantum phase
transitions (during calculation that $J_{2}=J_{1}=1$).

According to the DMRG method, the whole system is divided into system block A
and environment block B. For a quantum system containing two subsystems A and
B, the Von Neumann entropy\cite{57,58} is defined as the partial entanglement
entropy of two subsystems A or B\cite{59},%
\begin{equation}
S=S\left(  \rho_{\text{A}}\right)  =S\left(  \rho_{\text{B}}\right)
=-\text{Tr}\left(  \rho_{\text{A}}\log\rho_{\text{A}}\right)  =-\text{Tr}%
\left(  \rho_{\text{B}}\log\rho_{\text{B}}\right)  =-%
%TCIMACRO{\tsum \limits_{i}}%
%BeginExpansion
{\textstyle\sum\limits_{i}}
%EndExpansion
\lambda_{i}\log\lambda_{i}\text{,} \label{4}%
\end{equation}
where $\rho_{\text{A}}=$Tr$_{\text{B}}\left(  \rho_{\text{AB}}\right)  $,
$\rho_{\text{B}}=$Tr$_{\text{A}}\left(  \rho_{\text{AB}}\right)  $,
$\rho_{\text{AB}}=$ $\rho$ is the ground state density matrix of the system,
$\lambda_{i}$ is the nonzero eigenvalue of $\rho_{\text{A}}$ or $\rho
_{\text{B}}$.

\subsection{Energy density and entanglement entropy\label{sec4.1}}%

%TCIMACRO{\FRAME{ftbpFU}{5.7363in}{5.0185in}{0pt}{\Qcb{Ground state energy
%density $e_{0}$, energy gap $\Delta e_{1}$ and entanglement entropy $S$: (a)
%$e_{0}$ gradually decreases and converges to a stable value with the increase
%of $L$. (b) Variations of $\Delta e_{1}$ with $\gamma$ for $L$
%$=16,24,28,36,40$, where there is a small peak near $\gamma=0.54$ when $L=40$.
%(c) The second derivative of $e_{0}$ with respect to $\gamma$. (d) There are
%two peaks in the variations of $S$ with $\gamma$. The inset shows the first
%derivative of $S$ with respect to $\gamma$.}}{}{fig7.eps}%
%width 5.7363in;  height 5.0185in;  depth 0pt;  original-width 11.4155in;
%original-height 9.2985in;  cropleft "0";  croptop "1";  cropright "1";
%cropbottom "0";  filename 'Fig7.eps';file-properties "XNPEU";}}}%
%BeginExpansion
\begin{figure}
[ptb]
\begin{center}
\includegraphics[
height=5.0185in,
width=5.7363in
]%
{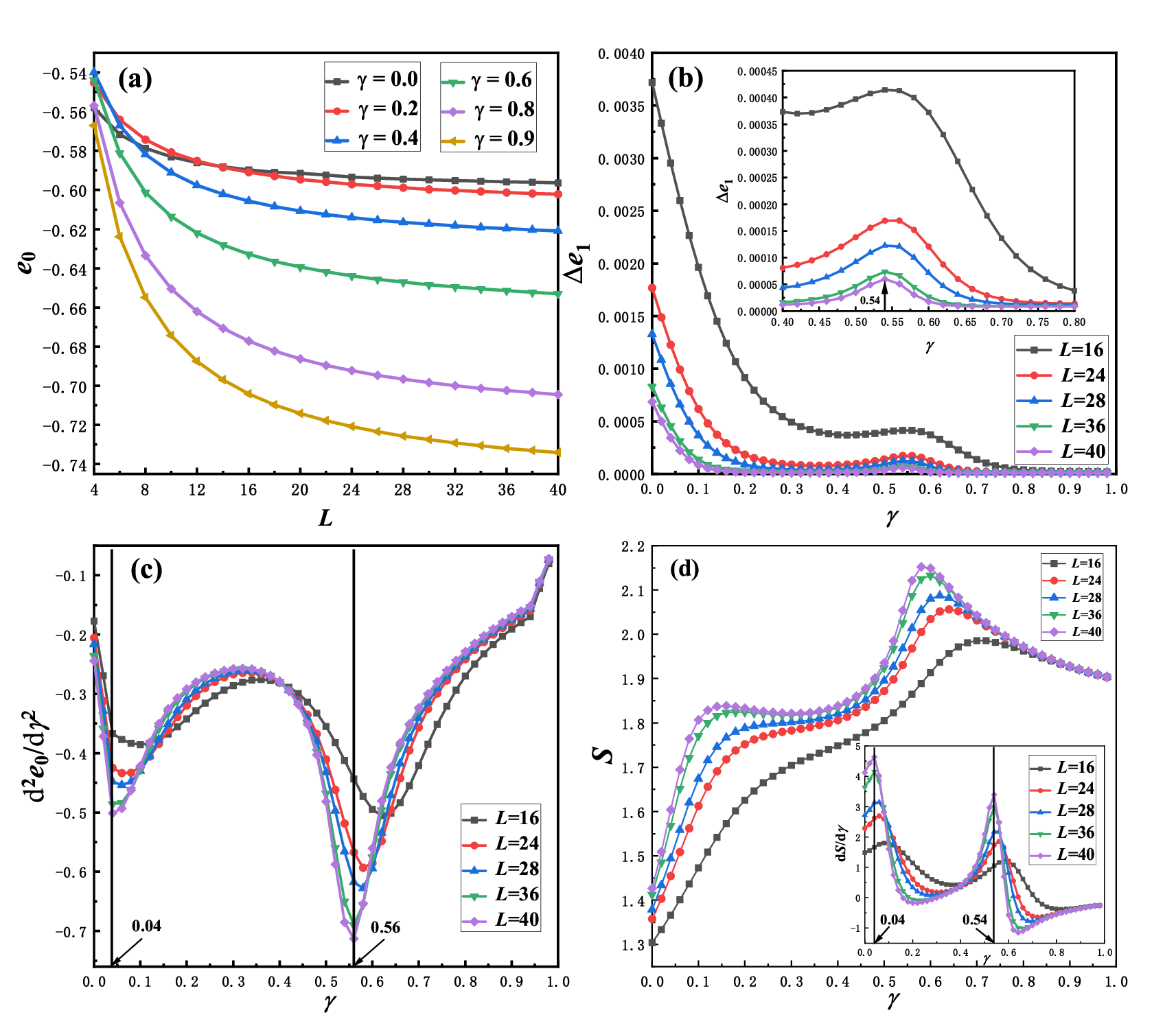}%
\caption{Ground state energy density $e_{0}$, energy gap $\Delta e_{1}$ and
entanglement entropy $S$: (a) $e_{0}$ gradually decreases and converges to a
stable value with the increase of $L$. (b) Variations of $\Delta e_{1}$ with
$\gamma$ for $L$ $=16,24,28,36,40$, where there is a small peak near
$\gamma=0.54$ when $L=40$. (c) The second derivative of $e_{0}$ with respect
to $\gamma$. (d) There are two peaks in the variations of $S$ with $\gamma$.
The inset shows the first derivative of $S$ with respect to $\gamma$.}%
\end{center}
\end{figure}
%EndExpansion

Without loss of generality, similar to the Sec. \ref{sec2}, we calculate the
energy density $e_{0}$, first excited state energy density $e_{1}$ and their
difference $\Delta e_{1}$, which vary with $L$ and $\gamma$ (see Fig. 7). From
Fig. 7(a), we find that $e_{0}$ gradually decreases and converges to a stable
value with the increase of $L$. Compared to the case when $\gamma=0$, the
convergence rate of $e_{0}$ to the stable value decreases with the increase of
$L$ due to the introduction of $\gamma$. For small values of $\gamma$, $e_{0}$
essentially converges when $L=28$, but for larger values of $\gamma$, the
convergence rate of $e_{0}$ is slower. As an example, we fit the data in Fig.
7(a) when $\gamma=0.9$ and get the relation between $e_{0}$ and $L$,
$e_{0}=0.358$exp$\left(  -L/2.715\right)  +0.124$exp$\left(  -L/12.352\right)
-0.738$. When $L\rightarrow\infty$, $e_{0}=$ $-0.738$. According to the data
of Fig. 7(a), when $\gamma=0.9$ and $L=28$, $e_{0}=-0.726$, which is near
$-0.738$. In order to study the phase transition, it is necessary to calculate
the case of $L\rightarrow\infty$. For convenience, the system with $L=28$ is
chosen to calculate the entanglement in the next subsection.

Next, we study the influence of $\gamma$ on the system energy and entanglement
entropy (see Figs. 7(b) - (d)). In Fig. 7(b), it is observed that when $L$
takes several values, $\Delta e_{1}$ first decreases with the increase of
$\gamma$, then slightly increases around $\gamma$ $=0.54$, and subsequently
drops to zero. Figure 7(c) indicates that with the increase of $\gamma$, there
are two minimum values of the second derivative of $e_{0}$ at $\gamma=0.04$
and $0.56$. Figure 7(d) shows that the $S$ varies with $\gamma$ when $L$ takes
different values, and the analysis reveals that there are two peaks of $S$
with the increase of $\gamma$, and the larger $L$ is, the more obvious the
peak values are. The inset in Fig. 7(d) shows that with the increasing $L$,
there are two maximum values of the derivative of $S$ at $\gamma=0.04$ and
$0.54$, respectively.

Based on the above analysis of energy density gap and entanglement entropy, we
predict that there are two phase transition points at about $\gamma=0.04$ and
$0.54$ in the system.%

%TCIMACRO{\FRAME{ftbpFU}{6.6227in}{2.1741in}{0pt}{\Qcb{Intraleg and interleg
%concurrence spatial distributions ($L=28,$ $J_{1}=J_{2}=1$). (a) and (b) are
%the spatial distributions of nearest-neighbour concurrence of chains-1 and 2
%when $\gamma=0,0.1,0.2,0.4,$ and the distributions of concurrence reverse the
%separation of odd and even bonds when $\gamma>0$, where the full-filled symbol
%represents concurrence on odd bonds, the half-filled symbol represents
%concurrence on even bonds. (c) is the spatial distribution of interleg
%concurrence between chains-1(3) and 2 when $\gamma=0,0.1,0.3,0.5,0.6,0.7$.}}%
%{}{fig8.eps}{\special{ language "Scientific Word";  type "GRAPHIC";
%width 6.6227in;  height 2.1741in;  depth 0pt;  original-width 11.4155in;
%original-height 9.9946in;  cropleft "0";  croptop "1";  cropright "1";
%cropbottom "0";  filename 'Fig8.eps';file-properties "XNPEU";}}}%
%BeginExpansion
\begin{figure}
[ptb]
\begin{center}
\includegraphics[
height=2.1741in,
width=6.6227in
]%
{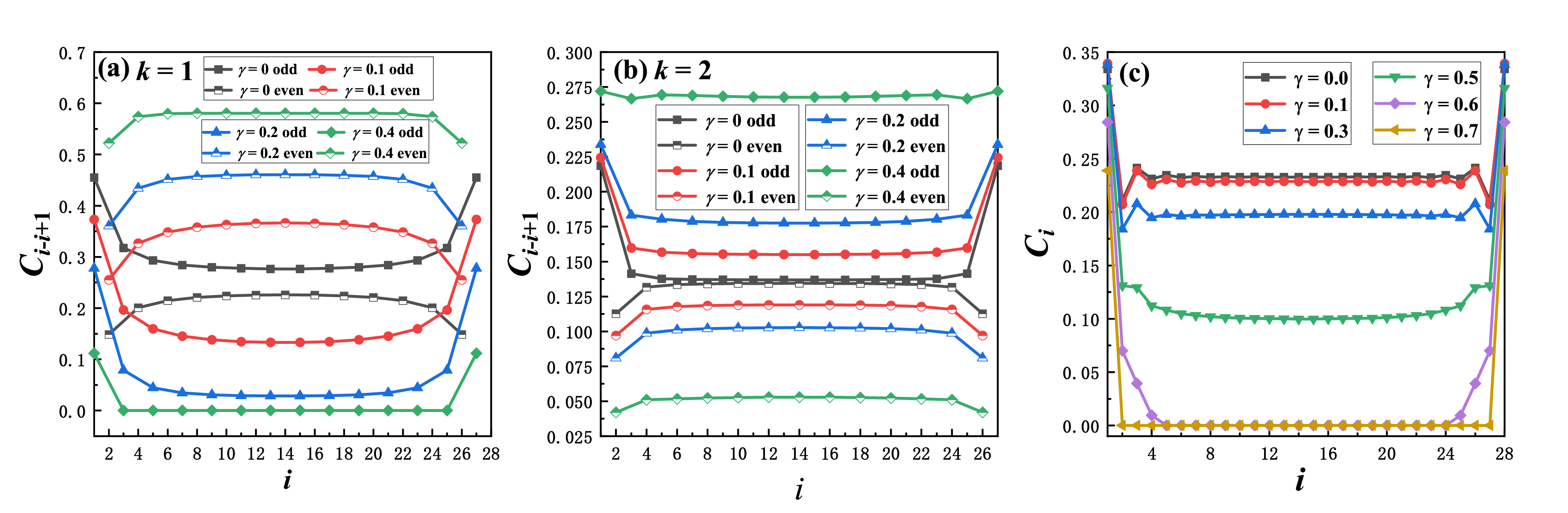}%
\caption{Intraleg and interleg concurrence spatial distributions ($L=28,$
$J_{1}=J_{2}=1$). (a) and (b) are the spatial distributions of
nearest-neighbour concurrence of chains-1 and 2 when $\gamma=0,0.1,0.2,0.4,$
and the distributions of concurrence reverse the separation of odd and even
bonds when $\gamma>0$, where the full-filled symbol represents concurrence on
odd bonds, the half-filled symbol represents concurrence on even bonds. (c) is
the spatial distribution of interleg concurrence between chains-1(3) and 2
when $\gamma=0,0.1,0.3,0.5,0.6,0.7$.}%
\end{center}
\end{figure}
%EndExpansion

\subsection{Effect of $\gamma$ on entanglement\label{sec4.2}}

This subsection investigates the effect of $\gamma$ on intraleg and interleg
concurrence by calculating the spatial distribution of the nearest-neighbour
spins concurrence in chain-1(3) and 2, along with the variations of $C_{1},$
$C_{2},$ $C_{13}$ and $C_{14}$ with $\gamma$ when $L=28$.

\subsubsection{spatial distributions of nearest-neighbor concurrence}

In Sec. \ref{sec3.1}, we have obtained that when $\gamma=$ $0$, due to the
influence of the OBC in the $x$ direction of the system, the phenomenon of odd
and even bond separation appears in the intraleg concurrence. Figures 8(a) and
(b) show the spatial distribution of nearest-neighbor entanglement
$C_{i\text{-}(i+1)}$ when $\gamma$ takes several values. It can be seen that
introducing $\gamma$ (when $J_{1}=J_{2}=1$, this problem can be seen as the
competition between $\gamma$ and OBC) makes the concurrence of odd and even
bonds in chain-1 gradually weaken and strengthen, respectively. In contrast to
the intraleg concurrence, the interleg concurrence distribution across the
rungs exhibits greater uniformity. Figure 8(c) shows that due to the influence
of boundary conditions, the interleg concurrence at both ends, $C_{1}$ and
$C_{28}$ is larger than $C_{14}$ at the middle of the ladder. In addition, the
concurrence values in the middle of the chain are nearly identical, indicating
that the dimerization effect brought by OBC only affects the concurrence of a
few bonds at both ends. As can be seen from Fig. 8(c), the interleg
concurrence has reached a uniform distribution in the middle of the ladder,
i.e., $C_{13}=C_{14}=C_{15}$.

\subsubsection{Further verification of phase transition point}

In order to further study the relation between quantum entanglement and phase
transitions of the system, we calculate the concurrence $C_{(L/2-1)\text{-}%
L/2}$ of the \textquotedblleft strong bonds\textquotedblright\ at the centre
of chain-2 (this selection is made to minimize the boundary effects), and
discuss the dependence of $C_{(L/2-1)\text{-}L/2}$ on $\gamma$ for different
system sizes $L$ (see Figs. 9(a) and (b)).%

%TCIMACRO{\FRAME{ftbpFU}{6.3529in}{2.175in}{0pt}{\Qcb{(a) Concurrence of the
%"strong bonds" $C_{(L/2-1)\text{-}L/2}$ in the middle of the chain varies with
%$\gamma$ for different system sizes $L$. (b) Variation of the derivative of
%$C_{(L/2-1)-L/2}$ with respect to $\gamma$ in chain-2. With the increase of
%$L$, the value of $\gamma$ gradually tends to $0.54$ when $dC_{(L/2-1)-L/2}%
%/d\gamma$ reaches the maximum value. (c) The relation between the concurrence
%of the four transverse bonds and the $\gamma$ for $L=28$. The inset shows the
%derivative of $C_{13}$ with respect to $\gamma$, where there is a minimum at
%$\gamma=0.54$.}}{}{fig9.eps}{\special{ language "Scientific Word";
%type "GRAPHIC";  maintain-aspect-ratio TRUE;  display "USEDEF";
%valid_file "F";  width 6.3529in;  height 2.175in;  depth 0pt;
%original-width 18.424in;  original-height 7.9649in;  cropleft "0";
%filename 'Fig9.eps';file-properties "XNPEU";}}}%
%BeginExpansion
\begin{figure}
[ptb]
\begin{center}
\includegraphics[
height=2.175in,
width=6.3529in
]%
{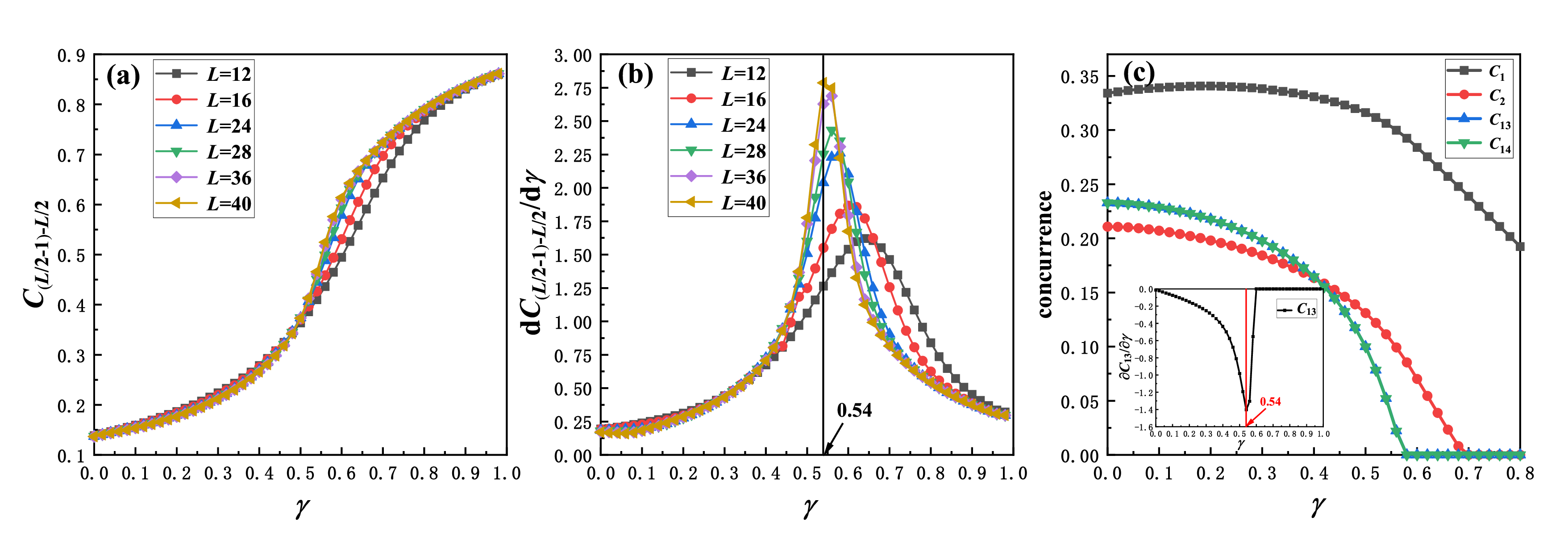}%
\caption{(a) Concurrence of the "strong bonds" $C_{(L/2-1)\text{-}L/2}$ in the
middle of the chain varies with $\gamma$ for different system sizes $L$. (b)
Variation of the derivative of $C_{(L/2-1)-L/2}$ with respect to $\gamma$ in
chain-2. With the increase of $L$, the value of $\gamma$ gradually tends to
$0.54$ when $dC_{(L/2-1)-L/2}/d\gamma$ reaches the maximum value. (c) The
relation between the concurrence of the four transverse bonds and the $\gamma$
for $L=28$. The inset shows the derivative of $C_{13}$ with respect to
$\gamma$, where there is a minimum at $\gamma=0.54$.}%
\end{center}
\end{figure}
%EndExpansion

Figure 9(a) shows the function of the $C_{(L/2-1)\text{-}L/2}$ in chain-2 with
$\gamma$ for different $L$\ values. The results indicate that the concurrence
monotonically increases with $\gamma$ and remains largely independent of the
system size $L$ (which means that the concurrence is insensitive to $L$) for
$\gamma$ $\leqslant0.5$ and $\gamma\geqslant0.9$; when $0.5<\gamma<0.9$, the
concurrence increases with $L$. A notable feature is the presence of a maximum
in the first derivative of $C_{(L/2-1)\text{-}L/2}$ with respect to $\gamma$,
as shown in Fig. 9 (b). It is important that as $L\rightarrow\infty$, the
extreme point $\gamma=0.54$ is close to the predicted phase transition point.

To further discuss the influence of $\gamma$ on interleg concurrence, the
$C_{1},$ $C_{2},$ $C_{13}$ and $C_{14}$ are studied for $L=28$ (see Fig.
9(c)). The results indicate that $C_{13}$ and $C_{14}$ gradually decrease to
zero at $\gamma=0.54$. In contrast, $C_{1}$ persists for a larger $\gamma$
value before diminishing to zero, a behaviour attributed to the dimerization
effect. The inset in Fig. 9(c) is the derivative of the middle bond
concurrence $C_{13}$($C_{14}$), in which there is a minimum value near
$\gamma=0.54$, which corresponds to the predicted phase transition point.

\subsection{Long-distance entanglement\label{sec4.3}}

Next, we investigate the LDE within the system. After calculation, one finds
that there are two types of LDE in this system, $C_{L}^{1}$ and $C_{L}^{2}$.
$C_{L}^{1}$ is the entanglement between sites $1$ and $L$ in $k=1(3)$, and
$C_{L}^{2}$ is the entanglement between site $1$ in $k=1(3)$ and site $L$ in
$k=3(1)$.

Figure 10(a) demonstrates that both $C_{L}^{1}$ and $C_{L}^{2}$ tend to be
stable with the increase of $L$ and converge completely at $L=28$, which
indicates that LDE is not sensitive to $L$. Consequently, high-intensity
entanglement can still be obtained in an infinite system. Comparing $C_{L}%
^{1}$ and $C_{L}^{2}$, one finds that when $L$ is very small, the two kinds of
LDE reach the same strength. It is indicated that for the ladder structure,
the difference between the two kinds of LDE gradually decreases to zero with
the increase of $L$, which can be easily seen from the ladder structure.%

%TCIMACRO{\FRAME{ftbpFU}{5.9516in}{2.7363in}{0pt}{\Qcb{(a) The LDE, $C_{L}^{1}$
%and $C_{L}^{2}$ versus $L$ for different values of $\gamma$, which reach the
%same intensity as $L$ is small. (b) $C_{L}^{1}$ and its derivative with
%respect to $\gamma$ for different values of $L$. The inset shows the variation
%of the derivative of $C_{L}^{1}$ with respect to $\gamma$, in which there is a
%maximum value$\ $at $\gamma=0.58$ as $L\rightarrow\infty$. }}{}{fig10.eps}%
%width 5.9516in;  height 2.7363in;  depth 0pt;  original-width 17.4208in;
%original-height 7.9649in;  cropleft "0";  croptop "1";  cropright "1";
%cropbottom "0";  filename 'Fig10.eps';file-properties "XNPEU";}}}%
%BeginExpansion
\begin{figure}
[ptb]
\begin{center}
\includegraphics[
height=2.7363in,
width=5.9516in
]%
{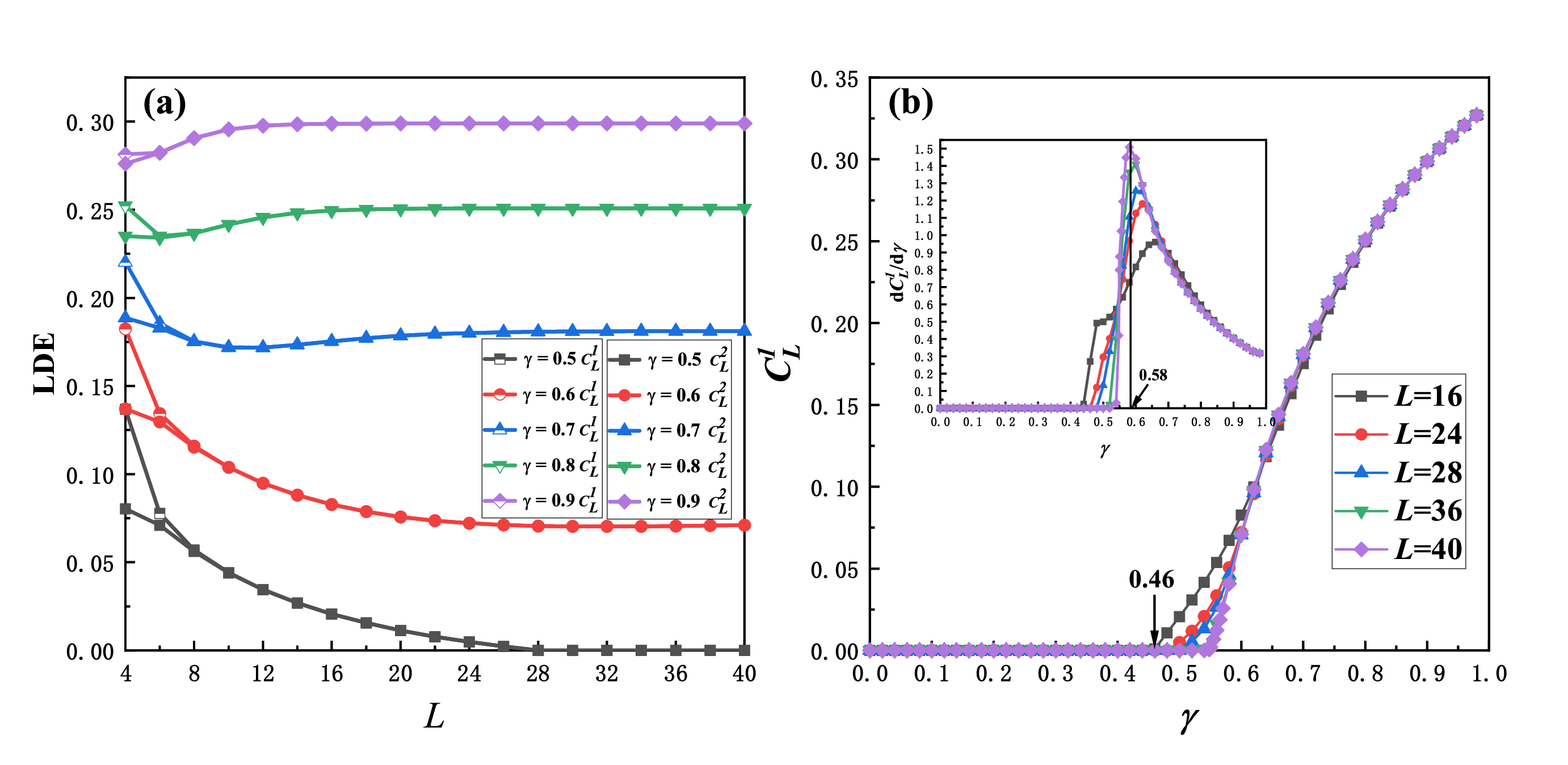}%
\caption{(a) The LDE, $C_{L}^{1}$ and $C_{L}^{2}$ versus $L$ for different
values of $\gamma$, which reach the same intensity as $L$ is small. (b)
$C_{L}^{1}$ and its derivative with respect to $\gamma$ for different values
of $L$. The inset shows the variation of the derivative of $C_{L}^{1}$ with
respect to $\gamma$, in which there is a maximum value$\ $at $\gamma=0.58$ as
$L\rightarrow\infty$. }%
\end{center}
\end{figure}
%EndExpansion

Figure 10(b) shows the variation of $C_{L}^{1}$ with $\gamma$ and the
convergence with the system size $L$. The result reveals that, due to the
influence of the interleg coupling $J_{2}$, the two-leg ladder requires a
larger $\gamma$ value to generate LDE compared to a single spin chain. Our
calculations indicate that the three-leg ladder necessitates a smaller
$\gamma$ value for LDE to generate in comparison to the two-leg ladder. Taking
$L=16$ as an example, the LDE in the two-leg ladder appears at about
$\gamma=0.51$, while $\gamma=0.46$ for the three-leg ladder. There is a
maximum value of the first derivative of $C_{L}^{1}$ with respect to $\gamma$
in the inset. As $L\rightarrow\infty$, the extreme point value $\gamma=0.58$
is near the critical point of the quantum phase transition.

\subsection{Effect of spin frustration on entanglement\label{sec4.4}}

As established in the previous section, for the system with the OBC, $\gamma$
can give rise to LDE. However, we find that LDE $=0$ in the system with CBC,
which can be interpreted as due to spin frustration that exists in the system
under CBC. Here, we will further study the effect of spin frustration in the
system on nearest-neighbor entanglement.%

%TCIMACRO{\FRAME{ftbpFU}{5.6092in}{2.7363in}{0pt}{\Qcb{Interleg concurrence
%between chains-1 and 3, $L=28$: (a) The spatial distribution of interleg
%concurrence. (b) The concurrence of the four interleg bonds varies with
%$\gamma$, $C_{1}$ and $C_{2}$ are generated when $\gamma=0.04$ and $0.12$,
%respectively, and $C_{13}$, $C_{14}$ are always zero.}}{}{fig11.eps}%
%width 5.6092in;  height 2.7363in;  depth 0pt;  original-width 17.4208in;
%original-height 7.9649in;  cropleft "0";  croptop "1";  cropright "1";
%cropbottom "0";  filename 'Fig11.eps';file-properties "XNPEU";}}}%
%BeginExpansion
\begin{figure}
[ptb]
\begin{center}
\includegraphics[
height=2.7363in,
width=5.6092in
]%
{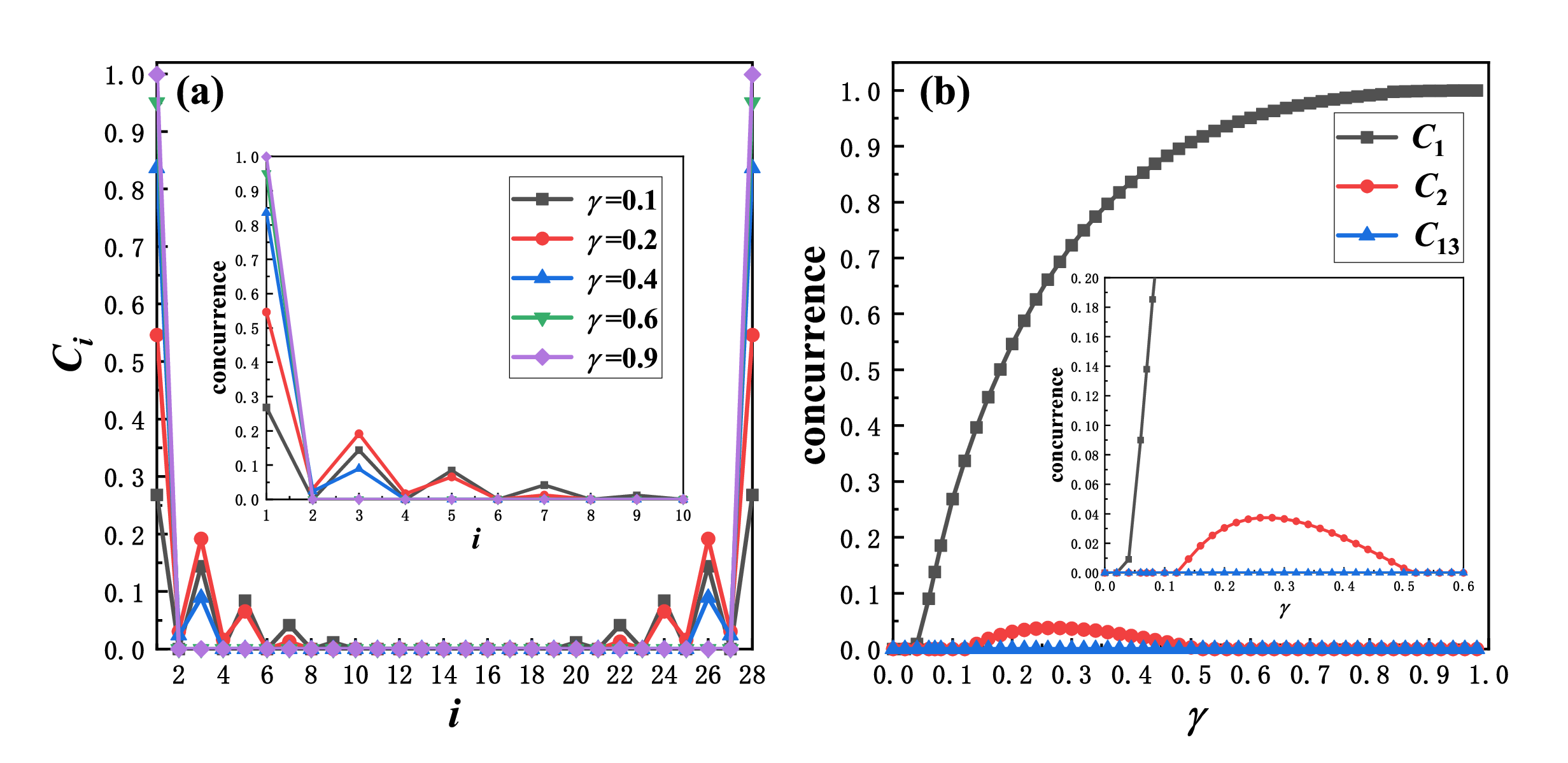}%
\caption{Interleg concurrence between chains-1 and 3, $L=28$: (a) The spatial
distribution of interleg concurrence. (b) The concurrence of the four interleg
bonds varies with $\gamma$, $C_{1}$ and $C_{2}$ are generated when
$\gamma=0.04$ and $0.12$, respectively, and $C_{13}$, $C_{14}$ are always
zero.}%
\end{center}
\end{figure}
%EndExpansion

In Sec. \ref{sec3.2}, we already know there is no interleg concurrence in the
system with $\gamma=0$ and CBC. However, the calculation in the case of CBC
shows that $\gamma$ can cause concurrence between chains-1 and 3. This
phenomenon arises because the distributions of intraleg spin interactions in
chains-1 and 3 are identical (see Fig. 1). As an example, we study the
distribution of the interleg concurrence $C_{i}$ and the variations of $C_{1}%
$, $C_{2}$, $C_{13}$ and $C_{14}$ with $\gamma$ (as shown in Fig. 11).

Due to the dimerization effect brought by OBC in the $x$ direction, $C_{1}$
and $C_{28}$ is the strongest at both ends of the ladder, and the concurrence
of odd and even bonds changes with the position $i$, and it decreases to zero
as $i\rightarrow14$. The even bond concurrence is very small and almost always
zero. As the increase of $\gamma$, except for the concurrence of the odd bonds
at the most ends of the ladder, the concurrence of the remaining bonds
eventually decreases to zero, and $C_{13}$ and $C_{14}$ of the middle interleg
bonds are always zero. For small values of $\gamma$, $C_{i}=0$, a consequence
of the suppressive influence of frustration on interleg concurrence. With the
increase of $\gamma$, $C_{1}$ is generated at $\gamma=0.04$ and then increases
rapidly. It is easy to see that $\gamma$ is conducive to the generation of
entanglement of odd bonds at the most ends of the ladder. Unlike $C_{1}$,
$C_{2}$ begins to appear at $\gamma=0.12$, increases and then decreases as
$\gamma$ increases, and disappears at $\gamma=0.52$.

\section{Conclusion\label{sec5}}

In this study, we investigate the spin-$1/2$ three-leg antiferromagnetic
Heisenberg ladder under OBC and CBC by calculating the dependence of the
energy density, entanglement entropy and concurrence on the interleg
interaction $J_{2}$ and the alternation parameter $\gamma$. In the case of
OBC, as $\gamma=0$, it is found that entanglement separation of odd and even
bonds occurs in the system. The introduction of $\gamma$ can completely
reverse the distribution of concurrence between odd and even bonds.
Additionally, $\gamma$ can induce the emergence of two types of LDE, i.e., the
intraleg LDE and the interleg one. The results show that the three-leg ladder
makes it easier to produce LDE than the two-leg system and makes the physical
mechanism and properties of LDE clearer, which is conducive to obtaining
stable entangled states effectively in experiments. We also find that the
nearest-neighbor spins entanglement, entanglement entropy, energy gap and LDE
are essentially related to phase transitions, and a phase transition point
near $\gamma=0.54$ is predicted by calculating.

Finally, the effect of CBC on entanglement is discussed, revealing that LDE
production is inhibited in the spin frustration. When $\gamma=0$, the interleg
concurrence is always in a dead state, and $\gamma$ can lead to interleg
concurrence appearing between chains-1 and 3.

\section{Acknowledgements}

This work is supported by the Shandong Provincial Natural Science Foundation,
China, under Grant No. ZR2022MA041, and No. ZR2021ME147; National Natural
Science Foundation of China under Grants No. 11675090, and No. 11905095;
Innovation Project for graduate students of Ludong University\ IPGS2024-049.


\begin{thebibliography}{99}                                                                                               %


\bibitem {1}A. Einstein, B. Podolsky, N. Rosen, Can quantum-mechanical
description of physicals reality be considered complete? Phys. Rev.
\textbf{47} (1935) 777.~

\bibitem {2}M.A. Nielsen, I.L. Chuang, Quantum Computation and Quantum
Information, Cambridge University Press, Cambridge, 2000.

\bibitem {3}R. Horodecki, P. Horodecki, M. Horodecki, K. Horodecki, Quantum
entanglement, Rev. Mod. Phys. \textbf{81} (2009) 865.

\bibitem {4}T. Nishioka, Entanglement entropy: Holography and renormalization
group, Rev. Mod. Phys. \textbf{90} (2018) 035007.

\bibitem {5}L. Amico, R. Fazio, A. Osterloh, V. Vedral, Entanglement in
many-body systems, Rev. Mod. Phys. \textbf{80} (2008) 517.~

\bibitem {6}D. Boschi, S. Branca, F. De Martini, L. Hardy, S. Popescu,
Experimental realization of teleporting an unknown pure quantum state via dual
classical and Einstein-Podolsky-Rosen channels, Phys. Rev. Lett. \textbf{80}
(1998) 1121.

\bibitem {7}N. Gisin, G. Ribordy, W. Tittel, H. Zbinden, Quantum cryptography,
Rev. Mod. Phys. \textbf{74} (2002) 145.

\bibitem {8}C.H. Bennett, G. Brassard, C. Cr\'{e}peau, R. Jozsa, A. Peres,
W.K. Wootters, Teleporting an unknown quantum state via dual classical and
Einstein-Podolsky-Rosen channels, Phys. Rev. Lett. \textbf{70} (1993) 1895.

\bibitem {9}O. Hirota, A.S. Holevo, C.M. Caves, Quantum Communication,
Computing, and Measurement, Plenum Press, New York, 1997.

\bibitem {10}A. Osterloh, L. Amico, G. Falci, R. Fazio, Scaling of
entanglement close to a quantum phase transition, Nature \textbf{416} (2002) 608-610.

\bibitem {10(1)}F.G.S.L. Brand\~{a}o, M. Horodecki, An area law for
entanglement from exponential decay of correlations, Nat. Phys. \textbf{9}
(2013) 721-726.

\bibitem {11}M. Kargarian, R. Jafari, A. Langari, Dzyaloshinskii-Moriya
interaction and anisotropy effects on the entanglement of the Heisenberg
model, Phys. Rev. A \textbf{79} (2009) 042319.

\bibitem {12}S.-J. Gu, S.-S. Deng, Y.-Q. Li, H.-Q. Lin, Entanglement and
quantum phase transition in the extended hubbard model, Phys. Rev. Lett.
\textbf{93} (2004) 086402.

\bibitem {13}J. Ren, Y.-M. Wang, W.-L. You, Quantum phase transitions in
spin-1 XXZ chains with rhombic single-ion anisotropy, Phys. Rev. A \textbf{97}
(2018) 042318.

\bibitem {14}F.-W. Ma, S.-X. Liu, X.-M. Kong, Quantum entanglement and quantum
phase transition in the XY model with staggered Dzyaloshinskii-Moriya
interaction, Phys. Rev. A \textbf{84 }(2011) 042302.

\bibitem {16}M. B. Hastings, Entropy and entanglement in quantum ground
states, Phys. Rev. B \textbf{76} (2007) 035114.

\bibitem {19(1)}T.J. Osborne, M.A. Nielsen, Entanglement in a simple quantum
phase transition, Phys. Rev. A \textbf{66} (2002) 032110.

\bibitem {19(2)}L.-A. Wu, M.S. Sarandy, D.A. Lidar, Quantum phase transitions
and bipartite entanglement, Phys. Rev. Lett. \textbf{93} (2004) 250404.

\bibitem {19(4)}T.R. de Oliveira, G. Rigolin, M.C. de Oliveira, E. Miranda,
Multipartite entanglement signature of quantum phase transitions, Phys. Rev.
Lett. \textbf{97} (2007) 170401.

\bibitem {19(6)}S.-J. Gu, H.-Q. Lin, Y.-Q. Li, Entanglement, quantum phase
transition, and scaling in the XXZ chain, Phys. Rev. A \textbf{68} (2003) 042330.

\bibitem {19(7)}J. Ren, X.-F. Xu, L.-P. Gu, J.-L. Li, Quantum information
analysis of quantum phase transitions in a one-dimensional V$_{1}$-V$_{2}$
hard-core-boson model, Phys. Rev. A \textbf{86} (2012) 064301.

\bibitem {20}M. Azuma, Z. Hiroi, M. Takano, K. Ishida, Y. Kitaoka, Observation
of a spin gap in SrCu$_{2}$O$_{3}$ comprising spin-1/2 quasi-1D two-leg
ladders, Phys. Rev. Lett. \textbf{73} (1994) 3463.

\bibitem {21}T. Imai, K.R. Thurber, K.M. Shen, A.W. Hunt, F.C. Chou, $^{17}$O
and $^{63}$Cu NMR in undoped and hole doped Cu$_{2}$O$_{3}$ two-leg spin
ladder A$_{14}$Cu$_{24}$O$_{41}$ (A$_{14}$=La$_{6}$Ca$_{8}$, Sr$_{14}$,
Sr$_{11}$Ca$_{3}$), Phys. Rev. Lett. \textbf{81} (1998) 220.

\bibitem {22}G. Chaboussant, Experimental phase diagram of Cu$_{2}$(C$_{5}%
$H$_{12}$N$_{2}$)$_{2}$Cl$_{4}$: A quasi-one-dimensional antiferromagnetic
spin-Heisenberg ladder, Phys. Rev. B \textbf{55} (1997) 3046.

\bibitem {22(1)}E. Dagotto, T.M. Rice, Surprises on the way from one- to
two-dimensional quantum magnets: The ladder materials, Science \textbf{271}
(1996) 618-623.

\bibitem {49}B. Frischmuth, S. Haas, G. Sierra, T.M. Rice, Low-energy
properties of antiferromagnetic spin-1/2 Heisenberg ladders with an odd number
of legs, Phys. Rev. B \textbf{55 }(1997) R3340.

\bibitem {23}D.C. Johnston, J.W. Johnson, D.P. Goshorn, A.J. Jacobson,
Magnetic susceptibility of (VO)$_{2}$P$_{2}$O$_{7}$: A one-dimensional
spin-1/2 Heisenberg antiferromagnet with a ladder spin configuration and a
singlet ground state, Phys. Rev. B \textbf{35} (1987) 219.

\bibitem {24}B. Frischmuth, B. Ammon, M. Troyer, Susceptibility and
low-temperature thermodynamics of spin-1/2 Heisenberg ladders, Phys. Rev. B
\textbf{54} (1996) R3714.

\bibitem {25}S. Fujiyama, M. Tikigawa, N. Motoyama, H. Eisaki, S. Uchida,
Nuclear spin relaxation in hole-doped two-leg ladders, J. Phys. Soc. Jpn.
\textbf{69} (2000) 1610-1613.

\bibitem {26}F. Neaf, X. Wang, Nuclear spin relaxation rates in two-leg spin
ladders, Phys. Rev. Lett. \textbf{84} (2000) 1320.

\bibitem {27}X.Q. Wang, L. Yu, Magnetic-field effects on two-leg Heisenberg
antiferromagnetic ladders: Thermodynamic properties, Phys. Rev. Lett.
\textbf{84} (2000) 5399.

\bibitem {28}J. Vannimenus,~G. Toulouse, Theory of the frustration effect. II.
Ising spins on a square lattice, J. Phys. C: Solid State Phys. \textbf{10}
(1977) L537.

\bibitem {29}D.S. Almeida, R.R. Montenegro-Filho, Quantum bicritical point and
phase separation in a frustrated Heisenberg ladder, Phys. Rev. B \textbf{108}
(2023) 224433.

\bibitem {30}S. Bose, Quantum communication through an unmodulated spin chain,
Phys. Rev. Lett. \textbf{91} (2003) 207901.

\bibitem {32}J. Kurmann, H. Thomas, G. M\"{u}ller, Antiferromagnetic
long-range order in the anisotropic quantum spin chain, Physica A \textbf{112}
(1982) 235-255.

\bibitem {33}L.C. Venuti, C.D.E. Boschi, M. Roncaglia, Long-distance
entanglement in spin systems, Phys. Rev. Lett. \textbf{96} (2006) 247206.

\bibitem {34}S. Sahling, G. Remenyi, C. Paulsen, P. Monceau, V. Saligrama, C.
Marin, A. Revcolevschi, L.P. Regnault, S. Raymond, J.E. Lorenzo, Experimental
realization of long-distance entanglement between spins in antiferromagnetic
quantum spin chains, Nat. Phys. \textbf{11} (2015) 255-260.

\bibitem {35}D.I. Bazhanov, I.N. Sivkov, V.S. Stepanyuk, Engineering of
entanglement and spin state transfer via quantum chains of atomic spins at
large separations, Sci. Rep. \textbf{8} (2018) 14118.

\bibitem {36}G. Gualdi, S.M. Giampaolo, F. Illuminati, Modular entanglement,
Phys. Rev. Lett. \textbf{106} (2011) 050501.

\bibitem {37}S.M. Giampaolo, F. Illuminati, Long-distance entanglement in
many-body atomic and optical systems, New J. Phys. \textbf{12} (2010) 025019.

\bibitem {38}J. Chen, K.-L. Yao, L.-J. Ding, Identification of intrinsic
gapped behavior in spin-1/2 ladder with staggered dimerization, Physica A
\textbf{391} (2012) 2306-2312.

\bibitem {39}X.-Y. Deng, L.-J. Kong, L. Qiang, String order and degenerate
entanglement spectrum of \textit{S }= 1/2 and \textit{S }= 1 Heisenberg
bond-alternating chains, Eur. Phys. J. B \textbf{87 }(2014) 247.

\bibitem {40}S. Wessel, S. Haas, Three-dimensional ordering in weakly coupled
antiferromagnetic ladders and chains, Phys. Rev. B \textbf{62} (2000) 316.

\bibitem {41}T. Kariyado, Y. Hatsugai, Topological order parameters of the
spin-1/2 dimerized Heisenberg ladder in magnetic field, Phys. Rev. B
\textbf{91} (2015) 214410.

\bibitem {42}C. Ding, Phase transitions and critical behaviors of XXZ ladders,
J. Stat. Mech. Theory Exp. \textbf{2020} (2020) 013102.

\bibitem {43}L.-Z. Hu, Y.-L. Xu, P.-P. Zhang, S.-W. Yan, X.-M. Kong,
Long-distance entanglement in antiferromagnetic XXZ spin chain with
alternating interactions, Physica A \textbf{607} (2022) 128170.

\bibitem {44}R.-X. Li, S.-L. Wang, Y. Ni, K.-L. Yao, H.-H. Fu, 0- and
2/3-magnetization plateaus in three-leg antiferromagnetic Heisenberg spin-1/2
ladders with leg-dimerization, Phys. Lett. A \textbf{378 }(2014) 970--974.

\bibitem {45}R.C. Al\'{e}io, M.L. Lyra, J. Stre\v{c}ka, Ground states,
magnetization plateaus and bipartite entanglement of frustrated spin-1/2
Ising-Heisenberg and Heisenberg triangular tubes, J. Magn. Magn. Mater.
\textbf{417} (2016) 294--301.

\bibitem {46}M. Greiter, H. Schmidt, Magnetic excitations in the site-centered
stripe phase: Spin-wave theory of coupled three-leg ladders, Phys. Rev. B
\textbf{83} (2011) 144422.

\bibitem {47}S.J. Gibson, R. Meyer, G.Y. Chitov, Numerical study of critical
properties and hidden orders in dimerized spin ladders, Phys. Rev. B
\textbf{83} (2011) 104423.

\bibitem {48}M. Azzouz, K. Shahin, G.Y. Chitov, Spin-Peierls instability in
the spin-1/2 Heisenberg three-leg ladder, Phys. Rev. B \textbf{76} (2007) 132410.

\bibitem {50}X. Wang, N. Zhu, C. Chen, Ground-state phase diagram of a
spin-1/2 frustrated three-leg antiferromagnetic Heisenberg ladder, Phys. Rev.
B \textbf{66} (2002) 172405.

\bibitem {51}S. Wang, S. Zhu, Y. Ni, L. Peng, R. Li, K. Yao, Quantum phase
transition and magnetic plateau in three-leg antiferromagnetic Heisenberg spin
ladder with unequal \textit{J}$_{1}$--\textit{J}$_{2}$--\textit{J}$_{1}$ legs,
J. Magn. Magn. Mater. \textbf{397} (2016) 319--324.

\bibitem {52}S.R. White, Density matrix formulation for quantum
renormalization groups, Phys. Rev. Lett. \textbf{69} (1992) 2863.

\bibitem {53}L. Hulth\'{e}n, \"{U}ber das austauschproblem eines kristalls,
Ark. Mat. Astron. Fysik, \textbf{26A} (1938) 1-106.

\bibitem {54}W.K. Wootters, Entanglement of formation of an arbitrary state of
two qubits, Phys. Rev. Lett. \textbf{80} (1998) 2245.

\bibitem {55}S.-W. Tsai, J.B. Marston, Density-matrix renormalization-group
analysis of quantum critical points: Quantum spin chains, Phys. Rev. B
\textbf{62} (2000) 5546.

\bibitem {56}N. Laflorencie, E.S. S\o ensen, M.-S. Chang, I. Affleck, Boundary
effects in the critical scaling of entanglement entropy in 1D systems, Phys.
Rev. Lett. \textbf{96} (2006) 100603.

\bibitem {57}J.v. Neumann, Mathematical foundations of quantum mechanics,
Princeton University Press, Princeton, 1996.

\bibitem {58}V. Vedral, M.B. Plenio, M.A. Rippin, P.L. Knight, Quantifying
entanglement, Phys. Rev. Lett. \textbf{78} (1997) 2275.

\bibitem {59}C.H. Bennett, H.J. Bernstein, S. Popescu, B. Schumacher,
Concentrating partial entanglement by local operations, Phys. Rev. A
\textbf{53} (1996) 2046.
\end{thebibliography}
\end{document}